\documentclass[twocolumn,aps,pra,showpacs,superscriptaddress,nofootinbib]{revtex4-2}

\usepackage[latin1]{inputenc}
\usepackage{graphicx,epstopdf}
\usepackage{amsmath}
\usepackage{hyperref}
\usepackage{amsfonts}
\usepackage{amsthm}
\usepackage{amssymb}
\usepackage{color}
\usepackage{latexsym}
\usepackage{times,txfonts}
\usepackage{graphicx}
\usepackage{flushend}
\usepackage{setspace}

\newcommand{\fig}[1]{{Fig.}}
\newcommand{\qc}[2]{quantum coherence}
\begin{document}

\title{Non-Markovianity through entropy-based quantum thermodynamics}

\author{J. M. Z. Choquehuanca}
\affiliation{Instituto de F\'isica, Universidade Federal Fluminense, Av. Gal. Milton Tavares de Souza s/n, Gragoat\'a, 24210-346, Niter\'oi, RJ, Brazil}

\author{F. M. de Paula}
\affiliation{Centro de Ci\^{e}ncias Naturais e Humanas, Universidade Federal do ABC, Avenida dos Estados 5001, 09210-580, Santo Andr\'e, SP, Brazil}

\author{M. S. Sarandy}
\affiliation{Instituto de F\'isica, Universidade Federal Fluminense, Av. Gal. Milton Tavares de Souza s/n, Gragoat\'a, 24210-346, Niter\'oi, RJ, Brazil}

\date{\today}

\begin{abstract}
We introduce a generalized approach to characterize the non-Markovianity of quantum dynamical maps via breakdown of monotonicity of thermodynamic functions. By adopting an entropy-based formulation of quantum thermodynamics, we use the relationship between heat and entropy to propose a measure of non-Markovianity based on the heat flow for single-qubit quantum evolutions. This measure can be applied for unital dynamical maps that do not invert the sign of the internal energy. Under certain conditions, it can also be extended for other thermodynamic functions, such as internal energy and work flows. In this context, a natural connection between heat and quantum coherence can be identified for dynamical maps that are both unital and incoherent. As applications, we explore dissipative and non-dissipative quantum dynamical processes, illustrating the compatibility between our thermodynamic quantifiers and the well-establish measure defined via quantum coherence. 
\end{abstract}


\maketitle

\section{Introduction}

Quantum thermodynamics~\cite{deffner,gemmer} is a research field in the making, with fruitful conceptual implications and a potential 
range of applications, mainly in the field of quantum technologies (see, e.g., Ref.~\cite{alexia}). It aims at generalizing the laws of 
thermodynamics to the quantum domain, rewriting thermodynamic state functions and processes from the bottom up so that the laws of thermodynamics 
can be consistently applied to small systems and scaled up to the classical world~\cite{kosloff,oppenheim, tien,xu,fernando,hoda}. Remarkably, the definition 
of thermodynamic variables, such as work and heat, has been challenging,   since they cannot be taken as observables described by Hermitian operators~\cite{peter}. 
In the standard framework of quantum thermodynamics~\cite{alicki}, the internal energy change due to state population rearrangement induced by the external environment 
has been defined as {\it heat}. On the other hand, the internal energy change due to energy gap variations induced by the coherent dynamics has been defined as {\it work}. 
This framework has recently been revisited through an entropy-based formulation for heat and work~\cite{alipour1,alipour2,Ahmadi2020}, with heat identified as the 
entropy-related contribution to the internal energy change and work taken as the part unrelated to the entropy change. In particular, this has been applied to the study 
of the thermodynamics of a single qubit far from equilibrium~\cite{vallejo}, where a notion of temperature compatible with the classical limit is introduced. 
The alternative entropy-based view turns out to motivate an interplay with quantum information, with the interpretation of thermodynamic variables in terms of resources 
and correlations in a quantum system. 
  
Indeed, quantum thermodynamics is intrinsically connected with quantum information theory, with quantum resources strongly affecting thermodynamic variables and processes. 
This is manifested in a number of situations, such as measurements and their influence on the work performed on/by a quantum system~\cite{peter,william, daniel,michal,chen}, 
efficiency of thermodynamic tasks in quantum correlated devices~\cite{tien,Zhang:07,Lutz:14,Latune:19}, coherence effects on work and 
heat~\cite{Scully:03,philip,Brandner:17,streltsov,bernardo,latune,yuhan,kenza}, among others. By looking at information-based quantifiers, 
it is known that quantum coherence plays a relevant role in the characterization of system-bath memory effects. In particular, quantum coherence can be associated 
with a measure of non-Markovianity for {\it incoherent} dynamical maps~\cite{cramer,titas,passo,wu}. 
Notice that a memoriless (Markovian) behavior is always an idealization, with non-Markovian quantum dynamics being non-negligible 
in a number of different scenarios (see, e.g., Refs.~\cite{Apollaro:11,Haikka:12,Chen:15}). From an applied point of view, non-Markovian
dynamics may be also a resource for certain quantum tasks~\cite{Aharonov:06,Bylicka:14}. 

In this work, we adopt the entropy-based formulation of quantum thermodynamics to introduce a measure of non-Markovianity based on the heat flow for single-qubit quantum evolutions. 
This measure can be applied for {\it unital} dynamical maps that do not invert the sign of the internal energy (energy sign-preserving maps), which is rooted in the idea that von Neumann entropy itself can be used to build a 
non-Markovianity measure for unital maps~\cite{haseli}. Under certain conditions, we can also extend this approach to other thermodynamic functions, such as internal energy and work flows. Then, a natural connection between heat and quantum coherence can be observed for dynamical maps that are both unital and incoherent. 
In particular, we show that heat and work are monotonically related to quantum coherence for non-dissipative processes. Consequently, both heat and work can witness non-Markovianity for \textit{non-dissipative unital maps} 
as well as \textit{non-dissipative incoherent maps}. Moreover, we establish and illustrate the conditions over the system purity and the Hamiltonian such that the thermodynamic variables can be used to provide non-Markovianity measures 
throughout the evolution in a decohering environment.

The paper is organized as follows. In Section II, we briefly introduce the entropy-based formulation of quantum thermodynamics, expressing the first law of thermodynamics and 
its corresponding thermodynamic variables for a single qubit system. In section III, we discuss the non-Markovian dynamics, where we propose measures of non-Markovianity via 
generalized sign-preserving functions. In the section IV, we consider applications in dissipative and non-dissipative evolutions, 
illustrating the use of thermodynamic variables as measures of non-Markovianity. In section V, we present our conclusions.

\section{Quantum thermodynamics}

According to the standard framework for quantum thermodynamics~\cite{alicki}, the internal energy $U$ of a system described by a density operator $\rho$ is provided  
by the expected value of its Hamiltonian $H$, i.e., $U=\mathrm{tr}\left[\rho H\right]$. In this formalism, the first law of thermodynamics emerges from an infinitesimal change 
in the internal energy $dU=\delta Q+\delta W$, with $\delta Q=\mathrm{tr}\left[d\rho H\right]$ and $\delta W=\mathrm{tr}\left[\rho dH\right]$ defining the heat absorbed by system and 
the work performed on system, respectively. Considering the density operator as expressed in its spectral decomposition, we have 
$\rho=\sum_k r_k \left|r_k\right\rangle\left\langle r_k\right|$, where $\left|r_k\right\rangle$ denotes an eigenvector of $\rho$ and $r_k$ the corresponding eigenvalue. Then, 
the thermodynamic quantities $U$, $\delta Q$, and $\delta W$ can be rewritten as
\begin{equation}\label{Eq:energy}
U=\sum_k r_k \left\langle r_k\right|H\left| r_k\right\rangle,
\end{equation}
\begin{equation}\label{Eq:heat}
\delta Q=\sum_k dr_k \left\langle r_k\right|H\left| r_k\right\rangle+\sum_k r_k \left(\left\langle r_k\right|H\,d\left|r_k\right\rangle+h.c.\right),
\end{equation}
\begin{equation}\label{Eq:work}
\delta W=\sum_k r_k \left\langle r_k\right|dH\left|r_k\right\rangle.
\end{equation}
Note that the first term on the right-hand-side of Eq.~(\ref{Eq:heat}) is the responsible for changes in the von Neumann entropy,  $S= -k_{B}\mathrm{tr}\left[\rho\mathrm{ln}\rho\right]$, since $dS= -k_{B}\sum_k  dr_k \mathrm{ln}r_k$. 
Therefore, in order to connect the heat flow with the entropy change as in classical thermodynamics, an entropy-based formulation of quantum thermodynamics has recently been introduced in Refs.~\cite{alipour1,alipour2,Ahmadi2020}. 
In this framework, heat and work are redefined through 
\begin{equation}\label{Eq:heat2}
\delta\mathbb{Q}=\delta Q -\delta \mathbb{W}^*
\end{equation}	
and
\begin{equation}\label{Eq:work2}
\delta \mathbb{W}=\delta W +\delta \mathbb{W}^*,
\end{equation}
where $\delta \mathbb{W}^*$ is an additional work contribution given by
\begin{equation}\label{Eq:workd*}
\delta \mathbb{W}^*=\sum_k r_k \left(\left\langle r_k\right|H\,d\left|r_k\right\rangle+h.c.\right).
\end{equation}
The work $\delta \mathbb{W}^*$ is related to the variation $d\left|r_k\right\rangle$ of the density operator eigenvectors. Notice that the entropy-based formalism satisfies the first law of thermodynamics, 
i.e., $dU=\delta \mathbb{Q}+\delta \mathbb{W}$, being then equivalent to the standard framework for $\delta \mathbb{W}^{*}=0$. Remarkably, it can be shown that the existence of quantum coherence in $\rho$ 
in the energy eigenbasis $\{\left|h_k\right\rangle\}$ is a necessary ingredient for a non-zero work $\delta\mathbb{W}^*$ if the energy eigenvectors are fixed (i.e., for  
$d\left|h_k\right\rangle= 0$, $\forall k$). Here, we will define coherence through the $l_1$-norm \cite{cramer} of $\rho$ in the energy eigenbasis, reading
\begin{equation}\label{Eq:coherence}
C(\rho)=\sum_{k\neq l}\left|\left\langle h_k\right|\rho\left|h_l\right\rangle\right|.
\end{equation}
Indeed, observe that, if $H$ is constant and $\rho$ and $H$ have a common basis of eigenvectors, then $d\left|h_k\right\rangle = d\left|r_k\right\rangle = 0$. This leads to $ \delta \mathbb{W}^*=0$. 
Moreover, since a common basis for $H$ and $\rho$ implies that $\rho$ is diagonal in the energy eigenbasis $\{\left|h_k\right\rangle\}$, we will have $C(\rho)=0$.

For an arbitrary  single qubit system, the density operator can be written in the Pauli basis $\{I,\vec{\sigma}\}$ as
\begin{equation}
\rho=\frac{1}{2} \left(I+\vec{r}\cdot\vec{\sigma}\right),
\end{equation}
where  $\vec{r}=(x,y,z)$ is the Bloch vector.
For the Hamiltonian, we have $H=-\vec{h}\cdot\vec{\sigma}$.
We observe that $\vec{r}=(x,y,z)$ can be interpreted as a classical magnetic dipole moment immersed in an external magnetic field $\vec{h}$. 
Indeed, the quantities in Eqs.~(\ref{Eq:energy}) - (\ref{Eq:coherence}) reduce to \cite{vallejo}:
\begin{equation}\label{Eq:energyq}
U=-\vec{h}\cdot\vec{r},
\end{equation}
\begin{equation}\label{Eq:heatq}
\delta Q=-\vec{h}\cdot d\vec{r},
\end{equation}
\begin{equation}\label{Eq:workq}
\delta W=-\vec{r}\cdot d\vec{h},
\end{equation}
\begin{equation}\label{Eq:heat2q} 
\delta\mathbb{Q}=U_rdr,
\end{equation}
\begin{equation}\label{Eq:work2q} 
\delta \mathbb{W}=rdU_r,
\end{equation}
\begin{equation}\label{Eq:work*q}
\delta \mathbb{W}^{*}=-rh\,\hat{h}\cdot d\hat{r},
\end{equation}
and
\begin{equation}\label{coherenceq}
C=r\sqrt{1-\frac{U_{r}^2}{h^2}},
\end{equation}
where $r\equiv |\vec{r}|$ is the purity, $U_r\equiv U/r$ is the internal energy per unit of purity, $h\equiv |\vec{h}|$ is the positive energy eigenvalue, $\hat{r}\equiv\vec{r}/r$ is 
the dipole direction, and $\hat{h}\equiv\vec{h}/h$ is the external field direction. Note that heat and work in the entropy-based qubit formalism, namely, $\delta\mathbb{Q}$ and $\delta\mathbb{W}$, 
are necessarily associated with changes in $r$ (and consequently in $S$) and $U_r$, respectively. On the other hand, the standard contributions for these thermodynamic quantities, 
$\delta Q$ and $\delta W$, are required to be associated with variations in $\vec{r}$ and $\vec{h}$, respectively. 

An interpretation for $\delta \mathbb{W}^*$ in terms of the behavior of $\vec{r}$ can be obtained through 
the relation between $\delta \mathbb{W}^*$ and $C(\rho)$. First, a change in the eigenvectors of $\rho$ is required for non-zero $\delta \mathbb{W}^*$. By imposing $d\hat{h}=0$ (i.e., a fixed energy eigenbasis), 
we can express $\delta \mathbb{W}^*$ in Eq.~(\ref{Eq:work*q}) as $\delta \mathbb{W}^*=hCd\theta$, where $hC=|\vec{r}\times\vec{h}|$ denotes the absolute value of the torque 
on $\vec{r}$ induced by $\vec{h}$ and $\theta=\text{arcos}(\hat{h}\cdot\hat {r})$ is the angle between $\vec{r}$ and $\vec{h}$~\cite{vallejo}. Therefore, $\delta \mathbb{W}^{*}$ is 
equivalent to the energy cost required to rotate a magnetic dipole moment immersed in an external magnetic field, being proportional to coherence. 
More generally, $\delta \mathbb{W}^{*}$ represents the departure from the quasistatic dynamics~\cite{alipour1,alipour2}.

\section{Characterizing non-Markovianity}
\label{sec:III}

Let us suppose an open-system dynamical evolution governed by a time-local master equation 
\begin{eqnarray} \label{Eq:mastereq}
&&\hspace{-0.2cm}\dot{\rho} = {\cal L}_t \, \rho(t)= -i\left[H(t),\rho(t)\right] \nonumber \\
&&\hspace{0.5cm}+\sum_i \gamma_i (t) \left( A_i(t)\rho(t) A^\dagger_i(t) - \frac{1}{2} \left\{A^\dagger_i(t) A_i(t),\rho(t)\right\}\right), 
\end{eqnarray}
where ${\cal L}_t$ is the time-dependent generator, $H(t)$ is the effective Hamiltonian of the system, $A_i(t)$ are the Lindblad operators, $\gamma_i(t)$ are the relaxation rates, and the dot symbol ``$\cdot$" denotes 
time derivative. 
By taking $\gamma_i(t) \ge 0$, ${\cal L}_t$ assumes the Lindblad form at each instant of time \cite{Gorini1976}. Consequently, the master equation solution $\rho(t)=\Phi_{t,\,\tau}\rho(\tau)$ is obtained through a completely positive trace-preserving (CPTP) map $\Phi_{t,\,\tau} = {\cal T} \exp \left(\int_{\tau}^{t} dt^\prime{\cal L}_{t^\prime} \right)$, with ${\cal T}$ representing the chronological time-ordering operator. In this case, the dynamical map $\Phi_{t,\,\tau}$ satisfies the divisibility condition, i.e.,  $\Phi_{t,\,\tau}=\Phi_{t,\,r}\Phi_{r,\,\tau}$ $(t \ge r \ge \tau \ge 0$), which characterizes the Markovianity of the 
dynamical evolution. On the other hand, for  $\gamma_i(t) < 0$, the corresponding dynamical map $\Phi_{t,\,\tau}$ may not be CPTP for
intermediate time intervals and the divisibility property of the overall CPTP dynamics is violated, which characterizes a non-Markovian behavior \cite{Breuer2009,Rivas2010,Chruscinski2014}. 

Now, assume $F_{\alpha}(t) = F_{\alpha}(\rho(t))$ represents an arbitrary monotonic function of $t$ under divisible dynamical maps, where $\alpha=+1$ and $\alpha=-1$ indicate increasing and decreasing behaviors, 
respectively. Then, a sign change in $\dot{F}_{\alpha}(t)$ works as a witness of non-Markovianity. From this breakdown of monotonicity, we propose a measure of non-Markovianity as 
\begin{equation}\label{Eq:NF}
N_{F_{\alpha}}[\Phi]=\text{max}_{\rho_0}\int_{\text{sgn}\dot{F}_{\alpha}=-\alpha}{\left|\dot{F}_{\alpha}(t)\right|dt}. 
\end{equation}
The maximization in Eq.~(\ref{Eq:NF}) is performed over all sets of possible initial states, $\rho_0$, and the integration  extends over all time intervals for which the sign of $\dot{F}_{\alpha}(t)$ is $\text{sgn}\dot{F}_{\alpha}=-\alpha$.  
If $\{(t_i^k, t_f^k)\}$ represents the set of all time intervals for which $\text{sgn}\dot{F_{\alpha}}=-\alpha$, then we can write
\begin{equation}\label{Eq:NF2}
N_{F_{\alpha}}[\Phi]=\text{max}_{\rho_0}\sum_{k\, :\, \text{sgn}\dot{F}_{\alpha}=-\alpha}\left|F_{\alpha}(t_f^k)-F_{\alpha}(t_i^k)\right|.
\end{equation}
Notice that $N_{F_{\alpha}}[\Phi]$ quantifies the departure from the map divisibility. It has been previously adopted for several functions $F_{\alpha}(t)$, such as in Refs.~{\cite{Breuer2009,Chruscinski2014}. 
Here, we use the terminology {\it measure} for $N_{F_{\alpha}}[\Phi]$ in the way it has been employed in the seminal work in Ref.~\cite{Breuer2009} and not as it is properly used in
resource theories (see, e.g., Refs.~\cite{Horodecki:09,streltsov}). 
The choice for $F_{\alpha}(t)$ will depend on the dynamical map. First, let us consider the case of operations over incoherent states. By choosing a fixed basis $\{|i\rangle\}$ in a $d$-dimensional 
Hilbert space, incoherent states are defined by density operators $\rho_{inc}$ that are diagonal when expressed in the basis $\{|i\rangle\}$, namely, $\rho_{inc}=\sum_i c_i \, |i\rangle\langle i|$, with $c_i$ denoting an  
arbitrary complex amplitude. Then, it follows the notion of an {\it incoherent map}, which is a dynamical map 
leading any incoherent state to another incoherent state. For the case of incoherent quantum operations, it can be shown that quantum coherence $C(\rho(t))$ can witness non-Markovianity through  $F_{-1}(t)=C(\rho(t))$~\cite{titas}. 
We can also consider the case of a {\it unital map}, which maps the identity operator to itself, $\Phi(I)=I$. In this case, it can be shown 
that the von Neumann entropy can witness non-Markovianity through $F_{+1}(t)=S(\rho(t))$~\cite{haseli}. 
An operational approach to determine whether or not a dynamical map is unital or incoherent can be established through the {\it{sufficient conditions}} obtained from Eq.~(\ref{Eq:mastereq}): (i) if $[A_i,A_i^{\dagger}]=0$ then $\Phi$ is a {\it{unital map}}; 
(ii) If $\left\langle h_n\right|A_i\left|h_k\right\rangle\left\langle h_k\right|A^{\dagger}_i\left|h_m\right\rangle=0$ for all $k$ and $n\neq m$ then $\Phi$ is an {\it{incoherent map}} in the energy eigenbasis $\{\left|h_k\right\rangle\}$~\cite{titas}. 
There are several well known quantum channels that are both incoherent and unital, such as phase flip, bit flip, bit-phase flip, among others~\cite{Nielsen-Book}. For further applications of Eq.~(\ref{Eq:NF2}) as a non-Markovianity measure and its relationship with correlation measures, see also Ref.~\cite{Paula2016}. Notice that, in this generalized approach, $\alpha$ may depend on the initial state, enabling the use of thermodynamic quantities such as internal energy, heat, and work to characterize non-Markovianity.

The use of the von Neumann entropy to witness non-Markovianity for unital maps suggests that heat flow, as defined by the entropy-based formulation of quantum thermodynamics, is also able to detect non-Markovian processes for unital maps. 
Indeed, the expression $\dot{\mathbb{Q}}=U_r\dot{r}$ obtained from Eq.~(\ref{Eq:heat2q}) reveals that heat is {\textit{monotonically}} related to the purity (consequently to the entropy) for single qubit energy sign-preserving dynamics i.e., quantum 
evolutions such that the internal energy $U$ is either a non-negative or a non-positive function of the time $t$. Notice that the purity does not increase under unital Markovian quantum processes \cite{Streltsov2018}. Then, we can establish the following measure of 
non-Markovianity:
\begin{itemize}
\item [(\textit{a})] \textit{$N_{\mathbb{Q}}[\Phi]$ is a measure of non-Markovianity if $\Phi$ is a single-qubit energy sign-preserving unital map.}
\end{itemize}
In this case, heat can witness non-Markovianity via $F_{\alpha}(t)=\mathbb{Q}(\rho(t))$ with $\alpha=\text{sgn}U\neq 0$.  As an illustration, the sign of $U$ does not change under \textit{isochoric} processes, with heat a monotonic function of the purity
\begin{equation}\label{Eq:isochoric}
\mathbb{Q}=\Delta U=U_r\left(r-r_0 \right)\,\,\,\,(\dot{U}_r=0),
\end{equation}
where $r_0$ represents the initial purity. Then, we can employ either $\mathbb{Q}$ or $U$ to quantify the degree of non-Markovianity for \textit{isochoric unital maps}. Other examples include \textit{non-dissipative}  processes, where heat and work are monotonically related to the quantum coherence,
\begin{equation}\label{Eq:nondissipative}
\mathbb{Q}=-\mathbb{W}=U\mathrm{ln}\sqrt{\frac{C^2+U^2/h^2}{C_{0}^2+U^2/h^2}}\,\,\,\,(\dot{U}=0),
\end{equation}
being $C_0$ the initial quantum coherence. Consequently, we can use either $\mathbb{Q}$ or $\mathbb{W}$ to characterize the non-Markovianity of \textit{non-dissipative unital maps} as well as \textit{non-dissipative incoherent maps}. 
Regardless of $\mathbb{Q}$, we can use $U$ and $\mathbb{W}$ for characterization of non-Markovianity if $\vec{h}(t)\cdot \vec{r}(t)$ and $U_r(t)$ are monotonic functions of $t$ for $\gamma_i (t)\geq 0$ ($\forall \, t\geq 0$), respectively. 
As a special case, let us provide sufficient conditions for the thermodynamical variables to witness non-Markovianity for a time-independent Hamiltonian. In this scenario, we take $H=\omega_0\sigma_z$ and 
denote $\vec{r}=[x(t),y(t),z(t)]$. Moreover, let us define $z_r\equiv z/r = \text{cos}\theta$. Then, it follows that
\begin{equation}\label{Eq:energy-z}
U(t)=\omega_0z(t),
\end{equation}
\begin{equation}\label{Eq:heatflow-z}
\dot{\mathbb{Q}}(t)=\omega_0 z_r(t)\dot{r}(t),
\end{equation}
\begin{equation}\label{Eq:workflow-z}
\dot{\mathbb{W}}(t)=\dot{\mathbb{W}}^*(t)=\omega_0r(t)\dot{z}_r(t).
\end{equation}
Hence, from Eq.~(\ref{Eq:heatflow-z}), we can establish that: 
\begin{itemize}
\item [(\textit{b})] \textit{$N_{\mathbb{Q}}[\Phi]$ is a measure of non-Markovianity for a time-independent Hamiltonian $H=\omega_0\sigma_z$ if $\Phi$ is a single-qubit unital map that does not invert the sign of $z(t)$.}
\end{itemize}
Furthermore, it follows from Eq.~(\ref{Eq:energy-z}) and Eq.~(\ref{Eq:workflow-z}) that $N_U$ and $N_{\mathbb{W}}$ are measures of non-Markovianity if $z(t)$ and $z_r(t)$  are monotonic functions of time for $\gamma_i (t)\geq 0$ ($\forall\, t \geq 0$), respectively.

\section{Applications}

\subsection{A dissipative quantum evolution}
Let us first consider  a dissipative single-qubit dynamics described by a time-local master equation given by Eq.(\ref{Eq:mastereq}), with $H=\omega_0\sigma_z$, $A_i =\delta_{1,i}\, \sigma_x$, and $\gamma_i=\delta_{1,i}\,\gamma$, such that $\omega_0>0$ and $\gamma>0$. Then
\begin{equation} \label{Eq:mastereq-markov}
    \dot{\rho}(t)=-i\omega_0\left[\sigma_z, \rho(t)\right]+\gamma
    \left[\sigma_{x}\rho(t)\sigma_{x}-\rho(t)\right].
\end{equation}
This master equation generates a Markovian quantum process $\rho(t)=\Phi_{M}\,\rho_0=(I+\vec{r}(t)\cdot\vec{\sigma})/2$, where $\vec{r}(t)=[x(t),y(t),z(t)]$, 
whose solution for the Bloch vector is given by
\begin{equation} 
x(t) =\frac{e^{-\gamma t}}{2 \omega}\left[\alpha_x e^{\omega t}+\beta_x e^{-\omega t} \right], \label{xx1}
\end{equation}
\begin{equation} 
y(t) =\frac{e^{-\gamma t}}{2 \omega}\left[\alpha_y e^{\omega t}+\beta_y e^{-\omega t} \right], \label{yy1}
\end{equation}
\begin{equation} 
z(t)= z_0 e^{-2\gamma t},  \label{Eq:BlochVectorExample1}
\end{equation} 
where
\begin{equation} 
\alpha_x= \omega x_0 + \gamma x_0 - 2 \omega_0 y_0,
\end{equation}
\begin{equation} 
\beta_x= \omega x_0 - \gamma x_0 + 2 \omega_0 y_0,
\end{equation}
\begin{equation} 
\alpha_y= \omega y_0 - \gamma y_0 + 2 \omega_0 x_0,
\end{equation}
\begin{equation} 
\beta_y= \omega y_0 + \gamma y_0 - 2 \omega_0 x_0,
\end{equation}
with 
\begin{equation} 
\omega=\sqrt{\gamma^2 - 4 \omega_0^2}
\end{equation}
and $\vec{r}_0=[x_0,y_0,z_0]$ denoting the initial state. The Markovian map $\Phi_M$ is both unital and incoherent, since the Lindblad operators $A_i =\delta_{1,i}\, \sigma_x$ satisfy the conditions (i) and (ii) described in Sec.~\ref{sec:III}. Moreover, $\Phi_M$ is also a map that preserves the sign of $z(t)$ (note that $\text{sgn}z(t)=\text{sgn}z_0$ $\forall\, t\geq 0$). Consequently, we can use the monotonicity of $\mathbb{Q}(t)$ or $C(t)$ as functions of $t$ to observe the Markovianity of Eq.~(\ref{Eq:mastereq-markov}). Then
\begin{equation}\label{Eq:NF1}
N_{F_{\alpha}}[\Phi_M]=0,
\end{equation}
where $F_{\alpha}= \mathbb{Q}$, with $\alpha=\text{sgn}z_0\neq 0$ or $F_{\alpha}=C$ with $\alpha=-1$. Indeed, Fig. \ref{fig:1} illustrates the sign preservation of $\dot{F}_{\alpha}(t)$ for the initial state $\vec{r}_0=[1/2,0,1/2]$, 
where the heat and coherence flows reduce to
\begin{equation}
\dot{\mathbb{Q}}(t)=\frac{\omega_0 \gamma \left[2\omega_0^2(1-\cosh{(2\omega t)}) -\omega^2 e^{-2\gamma t}\right]e^{-2 \gamma t}}{\omega^2 e^{-2 \gamma t}+\gamma^2 \cosh{(2 \omega t)}+\omega \gamma \sinh{(2\omega t)}-4 \omega_0^2},
\label{qq1}
\end{equation}
\begin{equation}
\dot{C}(t)=\frac{2\omega_0^2 \gamma e^{-\gamma t}\left(1-\cosh{2 \omega t}\right)}{\omega \sqrt{\gamma^2 \cosh{(2 \omega t)}+\omega \gamma \sinh{(2\omega t)}-4\omega_0^2}}.
\label{cc1}
\end{equation}

\begin{figure}[ht!]
\centering
\includegraphics[scale=0.32]{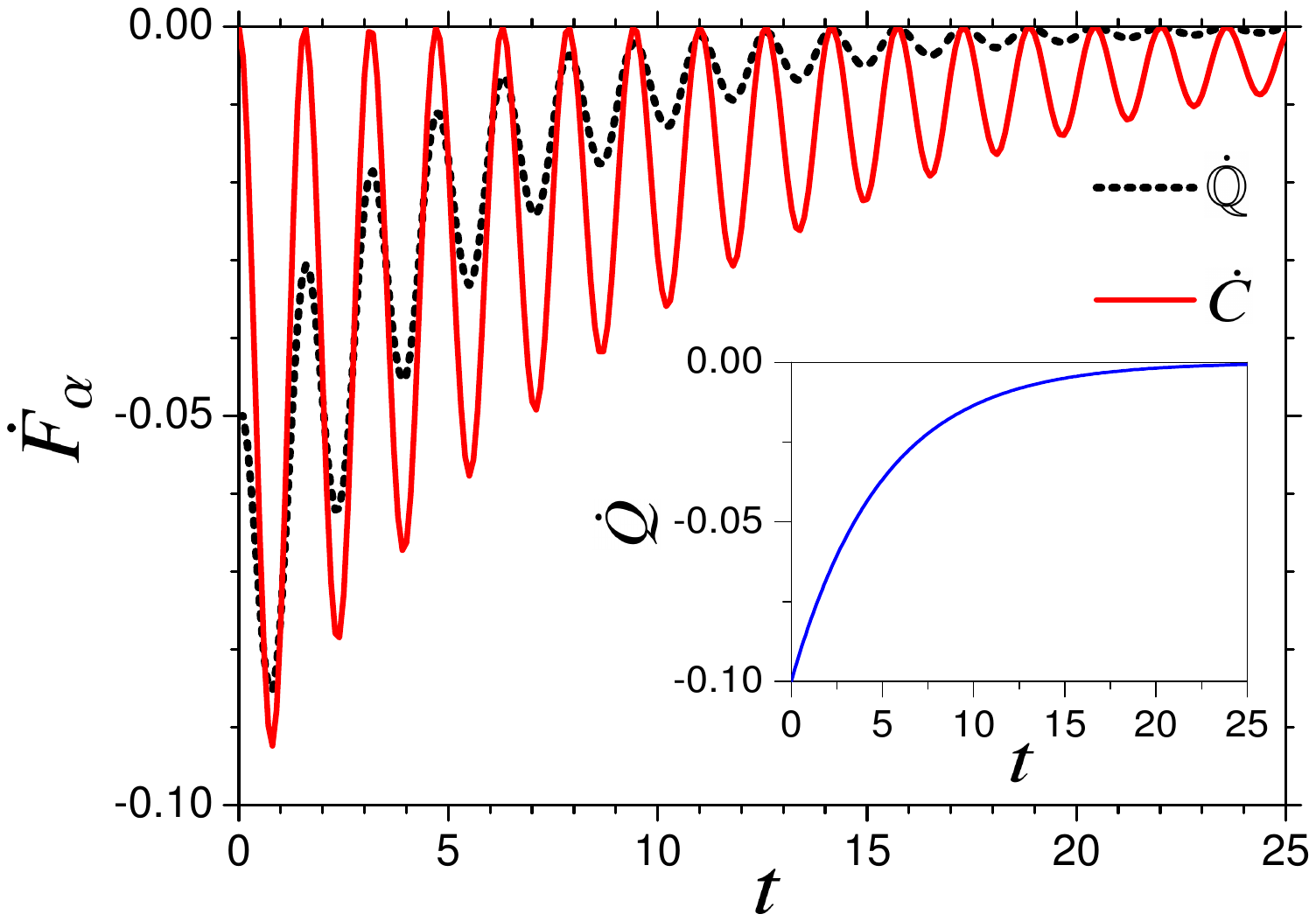}
\caption{(Color online) Heat (dotted black line) and quantum coherence (solid red line) flows as a function of time for $\vec{r}_0=[1/2,0,1/2]$, in units such that $\gamma=0.1$ and $\omega_0=1$. Inset: $\dot{Q}$ as a function of $t$.}
\label{fig:1}
\end{figure}

The Markovianity of Eq.~(\ref{Eq:mastereq-markov}) can also be witnessed from the behavior of the internal energy and work  (see plots for $U$ and $\mathbb{W}$ for $\vec{r}_0=[1/2,0,1/2]$ in Ref.~\cite{alipour1}). 
Concerning the standard thermodynamic quantities, we have that the conventional heat flow $\dot{Q}$ is equal to the internal energy flow $\dot{U}$. This holds for an arbitrary time-independent Hamiltonian, since 
the conventional work flow $\dot{W}$ is zero in this case. From Eqs.~(\ref{Eq:energy-z}) and~(\ref{Eq:BlochVectorExample1}), we then obtain $\dot{Q}=\dot{U}=-2\gamma z_0 \omega_0 e^{-2\gamma t}$. Since 
the map $\Phi_M$ preserves the sign of $\dot{Q}$ (see the inset plot in Fig. \ref{fig:1}), we can also conclude that $N_{F_{\alpha}}[\Phi_M]=0$ for $F_{\alpha}= Q$ with $\alpha=\text{sgn}z_0\neq 0$. 
From the inset in Fig. \ref{fig:1}, observe that the monotonicity between the heat flow and the coherence flow is lost in the conventional framework, but the identification of the Markovian behavior still works.  
In order to investigate a non-Markovian scenario, we may take temporary negative constant rates $\gamma$. In this case, Eqs.~(\ref{xx1}),~(\ref{yy1}), and (\ref{Eq:BlochVectorExample1}) 
are  kept for either positive or negative piecewise constant $\gamma$. 
This fact implies that the signs of internal energy, heat, and coherence flows will not be preserved throughout the dynamics, then identifying the non-Markovian behavior.  
Notice that the equivalence observed in this example between the 
conventional and entropy-based formalisms to characterize non-Markovianity is far from general. Indeed, we will consider next a non-dissipative quantum evolution for a time-indepedent Hamiltonian. 
As we will show for that case, both internal energy and conventional heat flow fail as non-Markovianity witnesses, since that $\dot{Q}=\dot{U}=0$ $\forall\, t\geq 0$. 

\subsection{A non-dissipative quantum evolution}

Let us consider now the dynamics of a single qubit under \textit{dephasing}, whose master equation is given by 
\begin{equation}\label{Eq:mastereq3} 
    \dot{\rho}(t)=-i\omega_0 \left[\sigma_z, \rho(t)\right]+\gamma(t)
    \left[\sigma_{z}\rho(t)\sigma_{z}-\rho(t)\right],
\end{equation}
This master equation can be derived from Eq.(\ref{Eq:mastereq}) by taking $H=\omega_0\sigma_z$, $A_i =\delta_{1,i}\, \sigma_z$, and $\gamma_i=\delta_{1,i}\,\gamma$. The solution is given by a map $\Phi_{D}$ such that $\rho(t)=\Phi_{D}\rho_0=(I+\vec{r}(t)\cdot\vec{\sigma})/2$ with \cite{titas}
\begin{equation} 
\vec{r}(t)=[x_0\Gamma (t), y_0\Gamma (t),z_0],
\end{equation}
where
\begin{equation} 
\Gamma(t)=\exp[-\int_0^t{\gamma(t) dt}].
\end{equation}
In this case, the internal energy and the quantum coherence are given by 
\begin{equation}\label{Eq:U-deph}
U(t)=U_0\, , \,\,\,\,\,\,\text{with}\,\,\,\,\,\,U_0=\omega_0z_0 , 
\end{equation}
and 
\begin{equation}\label{Eq:C-deph}
C(t)=C_0\Gamma (t)\, , \,\,\,\,\,\,\text{with}\,\,\,\,\,\,C_0=\sqrt{r_{0}^2-z_{0}^2}, 
\end{equation}
respectively. The  map $\Phi_D$ is both \textit{non-dissipative unital} and \textit{non-dissipative incoherent}. Consequently,  we can use $\mathbb{Q}(t)$, $\mathbb{W}(t)$, or $C(t)$ to 
characterize the non-Markovianity of Eq.(\ref{Eq:mastereq3}). Inserting Eqs.~(\ref{Eq:U-deph}) and (\ref{Eq:C-deph}) in Eq.~(\ref{Eq:nondissipative}), we can express heat in the form
\begin{equation}\label{Eq:Q-deph}
\mathbb{Q}(t)=\omega_0z_{r0}r_0\ln \sqrt{\Gamma(t)^2+\left(1-\Gamma(t)^2\right)z_{r0}^2}.
\end{equation}
The non-Markovianity measure in Eq.~(\ref{Eq:NF2}) then reads 
\begin{eqnarray} \label{Eq:Nmax}
&&\hspace{-0.5cm}N_{\mathbb{Q}}[\Phi_D]=\text{max}_{\rho_0}\sum_{k:\text{sgn}\dot{\mathbb{Q}}=\text{sgn}U_0}\left|\mathbb{Q}(t_f^k)-\mathbb{Q}(t_i^k)\right| \nonumber \\
&&\hspace{-0.5cm}=\omega_0\text{max}_{|z_{r0}|,r_0}r_0\sum_{k:\gamma < 0}|z_{r0}|\ln \sqrt{\frac{\Gamma(t_f^k)^{2}+(1-\Gamma(t_f^k)^{2})|z_{r0}|^2}{\Gamma(t_i^k)^{2}+(1-\Gamma(t_i^k)^{2})|z_{r0}|^2}} \nonumber \\
&&\hspace{-0.5cm}=\omega_0\text{max}_{|z_{0}|}\sum_{k:\gamma < 0}|z_{0}|\ln \sqrt{\frac{\Gamma(t_f^k)^{2}+(1-\Gamma(t_f^k)^{2})|z_{0}|^2}{\Gamma(t_i^k)^{2}+(1-\Gamma(t_i^k)^{2})|z_{0}|^2}}.
\end{eqnarray}
\begin{figure}[t]
\centering
\includegraphics[scale=0.3]{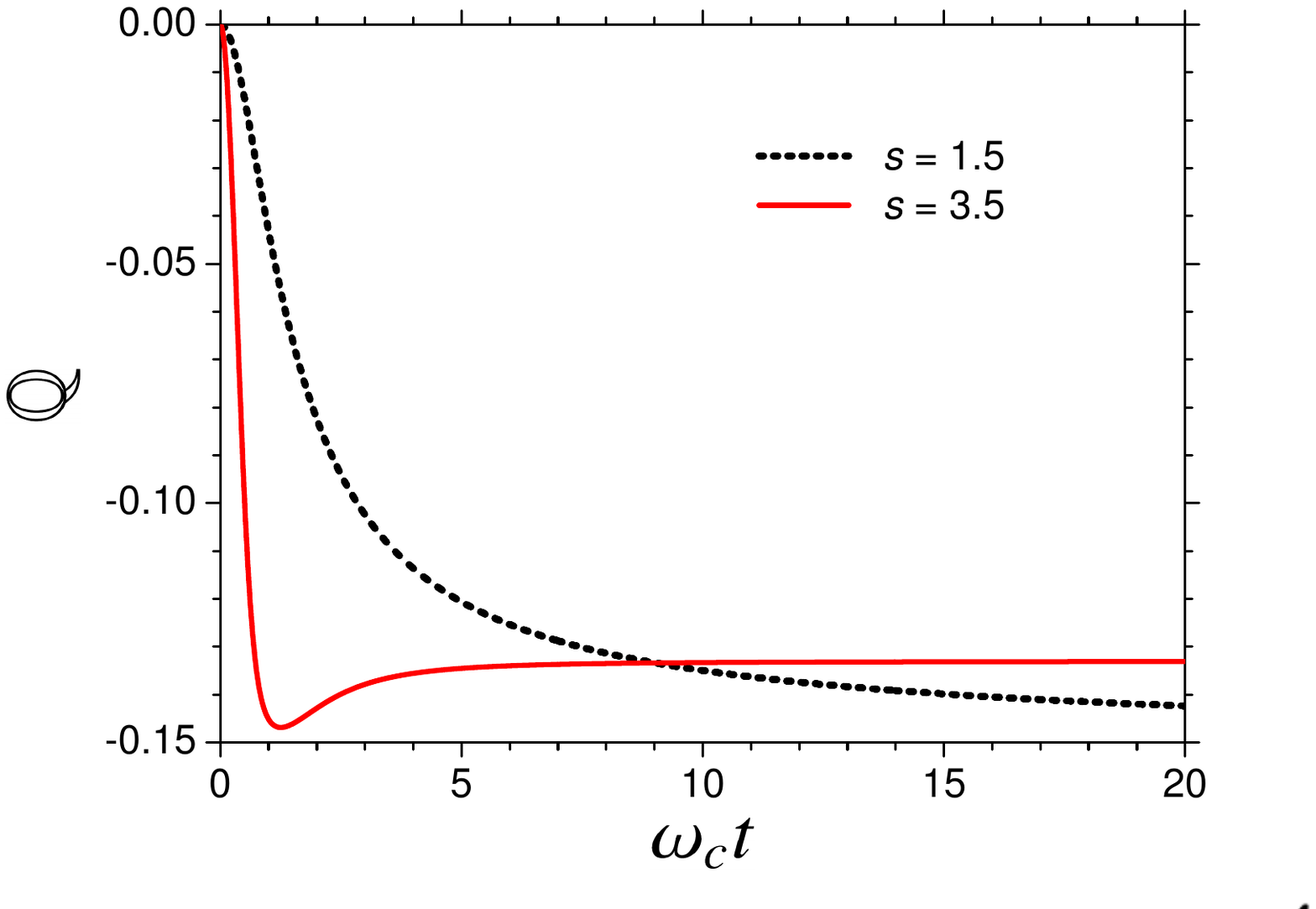}
\caption{(Color online)  $\mathbb{Q}$ as a function of $\omega_c t$ for $s = 1.5$ (dotted black line) and $s = 3.5$ (solid red line), with $\omega_0=1$, $r_0=1$, and $z_0=0.05$.}
\label{fig:2} 
\end{figure}
Note that a pure initial state (i.e., $r_0=1$) results from the maximization procedure in Eq.~(\ref{Eq:Nmax}). In order to enable a numerical comparison with the coherence-based
measure of non-Markovianity, let us consider a typical example of a zero-temperature bosonic reservoir with an Ohmic-like spectral density, where the time-dependent dephasing 
rate presents the specific form \cite{Haikka2013, Shekhar2015, titas}
\begin{equation}\label{Eq:gamma}
\gamma(t,s)=[1+(\omega_c t)^2]^{-s/2}\Gamma [s]\sin[s\arctan(\omega_c t)],
\end{equation}
being $s\geq 0$ the ohmicity parameter, $\Gamma[s]$ the Euler Gamma function, and $\omega_c$ the cutoff spectral frequency. In this case, $\gamma(t,s)<0$ occurs for 
$t_i^k=t_{2k-1}$ and $t_f^k=t_{2k}$, where $t_{k}=\tan[k\pi/2s]/\omega_c$, being $k$ an integer such that $0\leq k \leq \lfloor s\rfloor$ ($\lfloor s\rfloor$  is the floor function of $s$). 
Thus, we can conclude that $0\leq s \leq 2$ and $s> 2$ correspond to the Markovian and non-Markovian regimes of this model. In terms of the ohmicity parameter, $N_{\mathbb{Q}}[\Phi_D]$ takes the form
\begin{equation}\label{Eq:NQ-s}
N_{\mathbb{Q}}(s)=\omega_0\ln\left[\prod_{k=1}^{\lfloor s/2 \rfloor}\frac{\Gamma(t_{2k})^{2}+(1-\Gamma(t_{2k})^{2})z_{max}^2}{\Gamma(t_{2k-1})^{2}+(1-\Gamma(t_{2k-1})^{2})z_{max}^2}\right]^{\frac{z_{max}}{2}},
\end{equation}
where $z_{max}$ represents the value of $|z_0|$ that maximizes the last expression in Eq.~(\ref{Eq:Nmax}). The coherence-based measure $N_{C}[\Phi]$ has been employed 
to quantify the degree of non-Markovianity of the incoherent map $\Phi_D$ for the dephasing rate described in Eq.~(\ref{Eq:gamma}) \cite{titas}. As function of the $s$-parameter, $N_C[\Phi_D]$ reduce to
\begin{equation}\nonumber
N_{C}(s)=\text{max}_{\rho_0}\sum_{k:\text{sgn}\dot{C}=+1}\left|C(t_f^k)-C(t_i^k)\right|
\end{equation}
\begin{equation}\label{Eq:NC-s}
=\sum_{k=1}^{\lfloor s/2 \rfloor}\left[\Gamma(t_{2k})-\Gamma(t_{2k-1})\right],
\end{equation}
where a maximally coherent state (i.e., $C_0=1$) emerges from the maximization process. 

In order to witness non-Markovianity, Fig.~\ref{fig:2} shows the time evolution of the heat $\mathbb{Q}$ for an initial pure state under the single-qubit dephasing channel in both Markovian and non-Markovian regimes. 
Note that $\mathbb{Q}$ is a monotonically decreasing function of time for $s = 1.5$. On the other hand, a non-monotonic behavior arises due to the backflow of heat from the 
environment to the system for $s = 3.5$. The behaviors of the non-Markovianity measures $N_\mathbb{Q}$ and $N_C$ 
as functions of $s$ are illustrated in Fig.~\ref{fig:3}, with the inset showing $z_{max}$ versus $s$. These measures are monotonically related, 
i.e., $N_C(s)\approx 2.0 \, N_\mathbb{Q}(s)$, with both assuming non-zero values only for $s > 2$, a maximum value at $s=3.2$, and negligible values for $s > 5$. From the behavior of $z_{max}(s)$, 
notice that a maximally coherent state (where $z_0=0$) does not optimize the expression in Eq.~(\ref{Eq:NQ-s}) for all $s$.

\begin{figure}[t]
\centering
\includegraphics[scale=0.3]{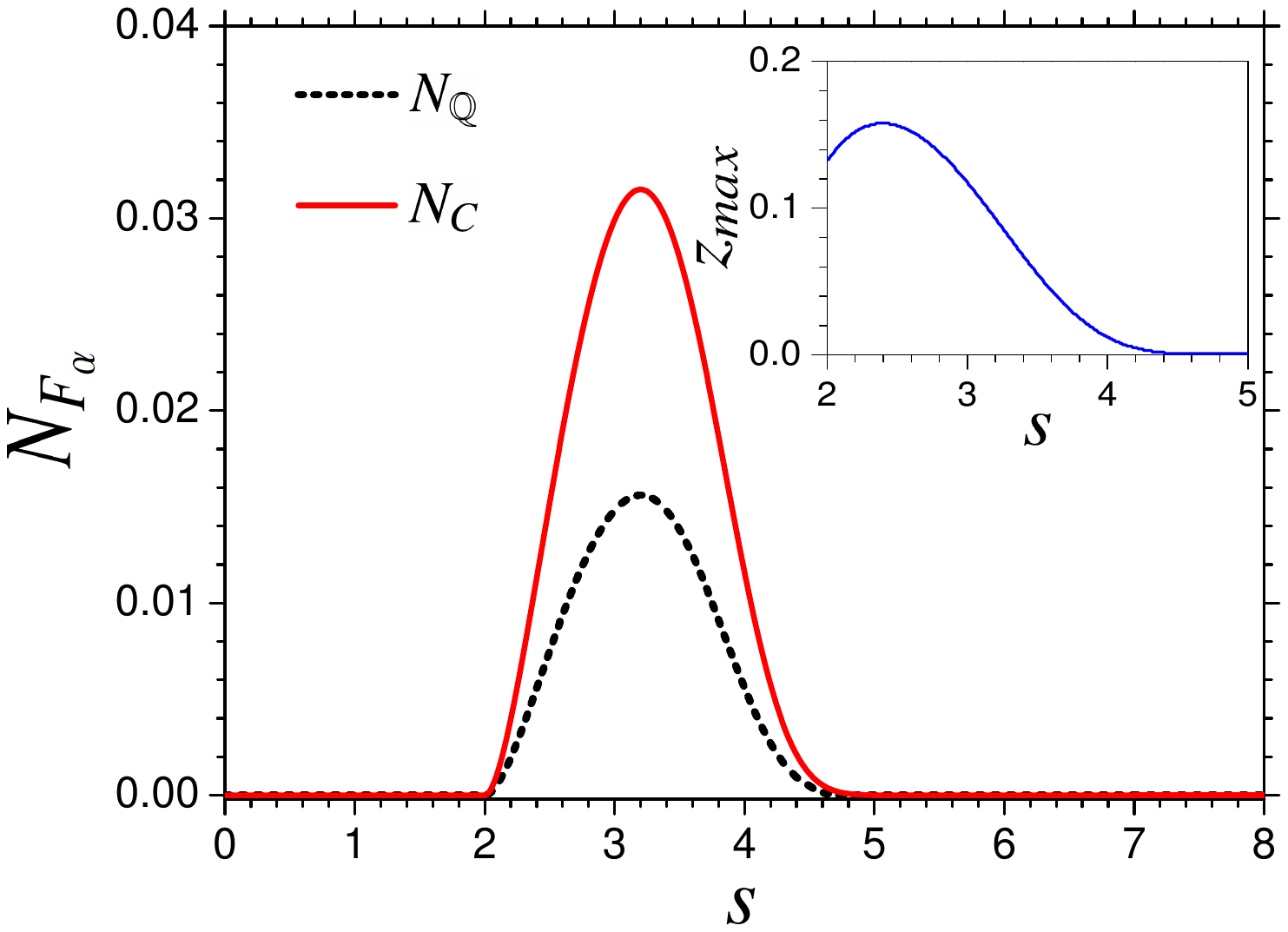}
\caption{(Color online) $N_\mathbb{Q}$ (dotted black line) and $N_C$ (solid red line) as functions of $s$ for $\omega_0=1$ and $\omega_c =1$. Inset: $z_{max}$ as a function of $s$.}
\label{fig:3}
\end{figure}

\section{Conclusions}

We have introduced a characterization of non-Markovianity by using the entropy-based formulation of quantum thermodynamics. 
Specifically, by looking at the divisibility property of the overall CPTP dynamics, we derived a measure of non-Markovian behavior 
through the relationship between entropy and heat in a thermodynamic process. This result supports the 
entropy-based approach, since it provides utility for the definition of heat by exclusively taking the entropy change contribution 
in the internal energy balance. 

The analysis of the heat flow can be a useful tool from an experimental point of view to provide a thermodynamic assessment of the memory effects of the bath. 
Based on the heat flow behavior, we can design an environment engineering approach to induce the desired amount of memory in a system-bath interaction 
for a suitable physical architecture. Moreover, from a conceptual point of view, it would be interesting to establish to what extent a thermodynamic characterization of non-Markovianity beyond single-qubit maps can be achieved. These are left as perspectives for future investigations. 


\section*{Acknowledgments}
J.M.Z.C. acknowledges financial support from  Conselho Nacional de Desenvolvimento Cient\'{\i}fico e Tecnol\'ogico (CNPq). 
M.S.S. is supported by Conselho Nacional de Desenvolvimento Cient\'{\i}fico e Tecnol\'ogico (CNPq) (307854/2020-5).
This research is also supported in part by Coordena\c{c}\~ao de Aperfei\c{c}oamento de Pessoal de N\'{\i}vel Superior (CAPES) (Finance Code 001) and by the Brazilian National Institute for Science and Technology of Quantum Information (INCT-IQ).


\end{document}